 \documentclass[final,5p,times,twocolumn,authoryear]{elsarticle}

\usepackage{amssymb}
\usepackage{lipsum}

\journal{Science of the Total Environment}

\usepackage{float}
\usepackage{enumitem}
\usepackage{textcomp}
\usepackage{mathtools}
\usepackage{commath}
\usepackage{float}
\usepackage{graphicx}
\usepackage{epstopdf}
\usepackage{caption}
\usepackage{hyperref} 
\usepackage{tocloft} 
\usepackage{fancyhdr} 
\usepackage{lastpage}
\usepackage{sidecap}
\usepackage[table,xcdraw]{xcolor}
\usepackage{subcaption}
\usepackage{indentfirst} 
\usepackage{tabularray}

\begin{document}

\begin{frontmatter}



\title{New approaches and error assessment to snow cover thickness and density using air temperature data at different heights.}

\author[UCM,IGEO]{Diego García-Maroto}
\author[UCM]{Luis Durán Montejano}
\author[Alcala]{Miguel Ángel de Pablo Hernández}

\affiliation[UCM]{organization={Departmento de Física de la Tierra y Astrofísica, Universidad Complutense de Madrid},
            city={Madrid},
            postcode={28040} ,
            country={España}}
\affiliation[IGEO]{organization={Instituto de Geociencias IGEO, CSIC-UCM},
            city={Madrid},
            postcode={28040} ,
            country={España}}
\affiliation[Alcala]{organization={Departmento de Geología, Geografía y Medio Ambiente, Universidad de Alcalá},
            city={Alcalá de Henares},
            postcode={28801}, 
            country={España}}

\begin{abstract}
Snow poles are inexpensive systems composed of a wooden mast with temperature sensors affixed at varying heights with the purpose of estimating the snow depth. They are frequently utilised in cold, remote regions where the maintenance of complex monitoring instruments becomes impractical. In this study, snow cover thickness is determined using different methods, based on the thermal behaviour of air temperature measured by a snow pole on Deception Island, Antarctica. The methods are compared to high-resolution measurements of snow depth obtained using an ultrasonic sensor at the same site. A new modified method is proposed and shown to give the best results. Errors and sensitivity to chosen thresholds of the various methods have been compared. Sensitivity tests have been also conducted to evaluate the impact of missing data from some of the sensors. Finally, the insulating effect on the thermal signal produced by the snow is used to obtain information on the snowpack density. Promising results have been found from this effort, opening new possibilities for the usage of snow poles and may lead to future studies. 

\end{abstract}



\begin{keyword}
Snow cover depth \sep Air temperature \sep Antarctica 



\end{keyword}

\end{frontmatter}

\thispagestyle{plain}


\section{Introduction}

The cryosphere is an essential part of the Earth's climate system. In particular, the snow cover plays a crucial role in shaping climate at high latitudes and in mountainous regions. Moreover, it can have a profound effect on the global climate through a phenomenon known as the snow-albedo feedback \citep{AR6_cryo}. The presence of a snowpack and its essential characteristics, including thickness and density, contribute significantly to the surface energy balance and to various ground-atmosphere exchanges. This is attributed to two distinct effects: firstly, the albedo of snow, which is much higher than that of most ground cover, reduces the amount of solar radiation absorbed by the ground, resulting in cooling \citep{albedo_feedback}. Secondly, due to its thermodynamic properties, including thermal conductivity, snow is a crucial component in the energy balance of the surface. The presence of a snowpack significantly reduces the energy exchange between the ground and the atmosphere by acting as a thermal insulator due to its low diffusivity \citep{judah_snow_climate}. These impacts have been extensively studied in polar and high mountain regions, as they can significantly affect the thermal regime of the ground surface \citep[e.g.][]{Ishikawa2003,dePablo2023} and contribute to changes in active layer thickness, temperature and permafrost state, even in places like Antarctica \citep[e.g.][]{dePablo2014,dePablo2017_ingles,ramos2017,Ramos2020}. 

As snow can impact both seasonal (active layer) and perennial (permafrost) frozen ground conditions \citep{Zhang2005}, monitoring its thickness is crucial in permafrost regions \citep{dePablo2016,dePablo2020}. Snow depth can be indirectly measured by using an array of temperature sensors that are attached to a wooden mast, commonly known as a ``snow pole''. These sensors record air temperatures at different heights above the ground, ranging such as 2.5, 5, 10, 20, 40, 80, and 160 cm \citep[e.g.][]{dePablo2014,dePablo2016,dePablo2020}. As one approaches greater heights above the ground, the separation between the sensors increases as, for the study of the effects of snow on the thermal regime of the permafrost, a better resolution is required closer to the ground \citep{dePablo2016}. This method exploits the insulating effect of the snow cover \citep{Zhang2005}, which leads to different daily temperature variations between sensors exposed to the atmosphere and those covered by the snow mantle \citep{Lewkowicz2008}. Although this approach does not result in an exact measurement of the snow's thickness, it provides an estimate based on the distance between consecutive sensors on the snow pole. However, this method is cost-effective: the sensors used, such as the Maxim iButton miniature temperature loggers \citep[e.g.][]{dePablo2016}, are relatively inexpensive, typically ranging from 50 to 100 euros. In contrast, conventional means, like snow pillows and ultrasonic devices, are inclined to be significantly more expensive \citep{alemanes}. For example, an ultrasonic sensor on its own may cost around 2000 euros, and a fully-equipped station dedicated to snow depth measurement can reach up to four times that, taking into account additional components such as the structure, data logger, batteries and other essential accessories.

Using specialised devices for measuring snow thickness can be too costly, especially when setting up multiple monitoring stations to examine the spatial variability of snow cover over large areas, as is often required for research into permafrost and active layer in polar regions. Furthermore, automated high-resolution equipment is prone to technical difficulties, such as power failures, resulting in prolonged data gaps. In contrast, should a miniature logger malfunction, the remaining loggers on the snow pole continue to autonomously acquire data, ensuring at least partial data availability \citep{dePablo2022,dePablo2023}. Additionally, the utilisation of miniature temperature loggers in snow poles proves cost-effective, even when numerous loggers are employed to achieve vertical resolution.

In this context, the method based on the semi-automatic analysis of air temperatures at different heights above ground \citep{danby, Lewkowicz2008} proves to be highly advantageous for several reasons: (1) it requires minimal logistical effort due to its simplistic installation, (2) it offers cost-effectiveness despite its lower resolution, and (3) it allows data collection to continue in the face of partial equipment failures, enabling research teams to establish more monitoring sites within their budgets. 

It is of interest to improve this method to reduce the resulting margin of error. Some researchers have attempted to improve the method by incorporating miniature temperature and light sensors \citep{paper_luz} into the snow poles, rather than relying on temperature data alone. However, such approaches have not been widely adopted among other research groups. This study therefore revisits the original method \citep{ Lewkowicz2008} and aims to achieve the following objectives:

\begin{itemize}
    \item Evaluate different methods from the existing literature for estimating seasonal snowpack thickness using an array of temperature sensors at different heights and compare these methods with reference data from an ultrasonic sensor.
    \item Develop a technique that integrates novel approaches with those from the literature to provide a more accurate estimate of snowpack thickness.
    \item Assess the potential errors and primary challenges associated with these methods.
    \item Obtain new ideas and approaches for future developments. 
\end{itemize}

Secondary objectives involve an exploration of whether air temperature data collected at the snow poles can provide additional information about the snowpack beyond its thickness. For example, it may be feasible to retrieve data on the thermal diffusivity from the effects that the snow produces on the thermal signal. Ultimately, this could provide insight into snow density and even snow water equivalent (SWE). This preliminary analysis is expected to guide further research.

\begin{figure*}[h!]
	\centering
	\includegraphics[width=0.95\textwidth]{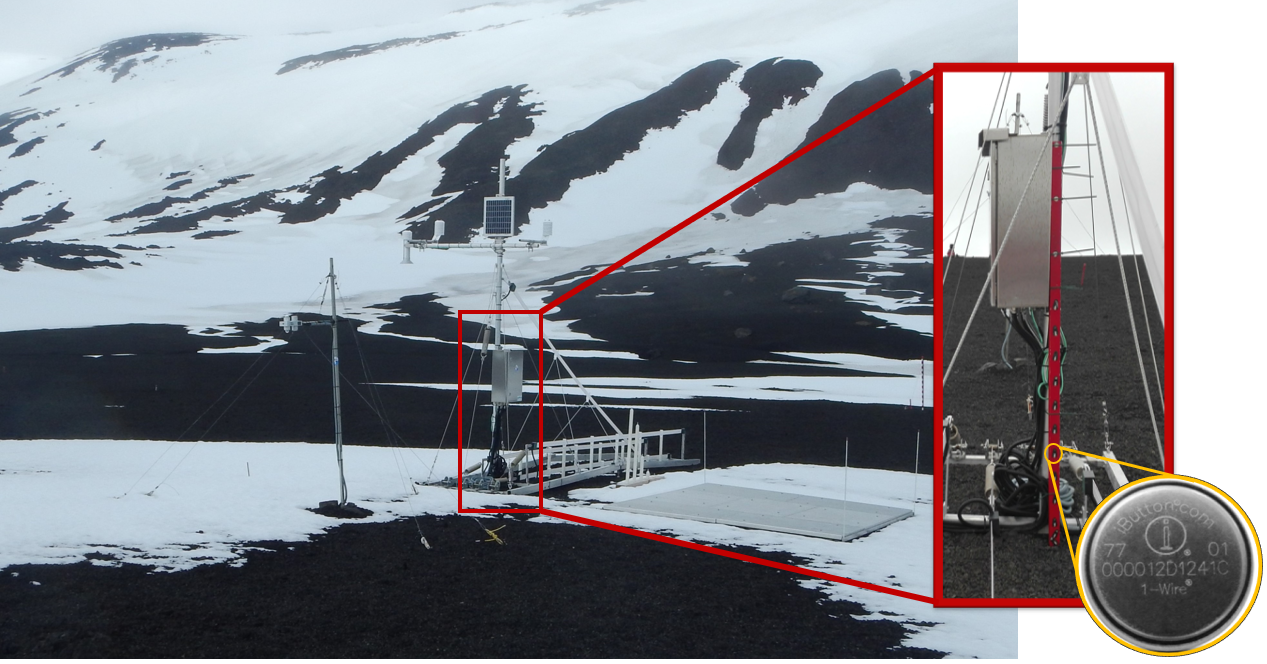}
	\caption{Picture of the Snow Pack Analyzer station at Crater Lake, Deception Island, Antarctica. Attached to the station’s main mast is the wood mast of the snow pole (detail in red box), including the array of miniature temperature logger -iButtons- (detail in yellow circle).}
	\label{fig: SPA_photo}
\end{figure*}

\section{Data}
\label{sec: data}
To achieve the main objectives, data collected between 2017 to early 2023 from a snow pole and an automatic snowpack analyser (SPA) station at the Crater Lake permafrost monitoring site on Deception Island in the South Shetland Archipelago, Antarctica \citep{dePablo2016, dePablo2020} (Fig. \ref{fig: SPA_photo}) was used. 

\subsection{Snow pole temperature data}
The data to be analysed hereafter consists of air temperature records taken every 3 hours at different heights above the ground, using iButton DS1922L miniature integrated temperature loggers from Maxim company, which have a resolution of 0.0625 °C and an accuracy of ±0.5 °C. These devices include the sensor, logger, and battery, and operate at a broad temperature range (-40 °C to 85 °C). While it is typical to install only a limited number of devices (usually 7) at the snow poles (de Pablo et al., 2014, 2021, 2022), this study utilized 15 loggers affixed to a wooden mast (snow pole) at heights of 2.5, 5, 10, 20, 30, 40, 50, 60, 70, 80, 90, 100, 120, 140 and 160 cm above the ground (Fig. \ref{fig: SPA_photo}). The use of a wooden mast is more appropriate than a metal one due to its lower thermal transmittance. This ensures (1) independent temperature records at different sensors and (2) low thermal irradiation to the snowpack from the mast when it is heated by the sun. Moreover, to prevent direct solar heating of the devices, they were installed on the north-facing side (shaded areas in Antarctica).

\subsection{Snow cover depth reference data}
In order to verify the results obtained when calculating snow depth based on temperature data, accurate hourly data on snowpack thickness were used. This data was collected from a SPA station \citep{dePablo2016,dePablo2020}, located in the same area of Deception Island (see Figure 1). The station records several environmental and snow cover variables on an hourly basis (see Table 1). These include thickness, density, ice and water content, and snow water equivalent (SWE) at various heights as well as for the entire snowpack. Additionally, the SPA features sensors to measure air temperature, moisture, solar direct and reflected radiation, and an ultrasonic sensor (Sommer USH-9), which provides a 1 mm resolution measurement of snow cover thickness with an accuracy of 0.1\% (refer to Figure 1). Data is collected by an MD8 logger installed at the station and powered by a 30W solar panel and a 12V 15Ah battery.

\subsection{Data loss and instrumentation failures}
The field data were recorded between 2017 and early 2023. However, different problems in the data and/or the devices and sensors have been reported along this period. 

Due to their limited data storage memory, the miniature loggers on the snow poles have to be retrieved and brought back to base for data extraction. Consequently, the data recorded during this time frame may be missing or need to be discarded, as it represents the temperature inside the Antarctic station. Nonetheless, this typically does not pose a significant issue as this process is carried out during austral summer, when there is little to no snow, so in most cases no relevant data is lost. Most of the databases utilised in this study have already undergone processing and validation. However, some additional post processing and validation is still required. Concerning the snow pole data, out of the 15 sensors installed, only 13 or 14 are active in most years as some sensors may fail to record data throughout the entire period. This could potentially impact the accuracy of the estimated snow depth calculation but does not signify a total data loss. Another sporadic issue with these sensors is the recording of erroneous data in the form of sudden jumps to impossible values for the climatology of Deception Island. These values can be easily detected and removed through visual inspection or application of impossible value or impossible increment tests. Other significant problems involve the extreme conditions of the Antarctic regions. For example, in 2017,  the snow pole data were rendered useless due to the pole being knocked down by strong wind gusts. In an effort to prevent this issue in the future, the snow pole was attached to the main mast of the SPA station the following year (as depicted in Figure \ref{fig: SPA_photo}). However, in 2020, the winds were strong enough to completely knock down the entire SPA station, rendering months of data from both the station and the snow pole useless.

The primary issue with the SPA data series is the presence of missing values, mainly due to power failures. Such an occurrence is severe, as it may result in the complete loss of data over several months, even jeopardising entire snow seasons. For example, in 2022, the station shut down in late May due to a failure in the main battery, and, as the sites are only visited during Antarctic summer field campaigns, the problem could not be resolved until January 2023. 

Due to the aforementioned issues, this study uses the 2019 series for analysis, as it provides the most complete set of data, with almost no relevant SPA station failures and all 15 snow pole sensors operating trouble-free. Other years, such as 2018, which had some temperature logger failures but complete SPA series data, are also employed to validate the developed algorithms.

\begin{table}[h!]
\caption{Parameters (and units) measured by the different instruments that integrate the Snow Pack Analyzer station in Deception Island. Note: parameters signaled with * are measured both for the whole snowpack and at specific heights above the ground (10, 30 and 50 cm) (modified from \citep{dePablo2020}).}
\label{tab: SPA_data_tab}
\centering
\resizebox{1\linewidth}{!}{%
\begin{tblr}{
  hline{1-2,10} = {-}{},
}
\textbf{Enviromental}   & \textbf{Thermal}                & \textbf{Snow properties}                               \\
Date @ Time             & Device temperature (°C)         & Ice content (\%)*                                      \\
Air temperature (°C)    & Surface temperature (°C)        & Water content (\%)*                                    \\
Air humidity (\%)       & Ground surface temperature (°C) & Density (kg/m\textsuperscript{3})*    \\
Pyranometer Up (W/m2)   & Temperature 10 cm (°C)          & SWE (mm)*                                              \\
Pyranometer Down (W/m2) & Temperature 20 cm (°C)          & Snow weight (kg/m\textsuperscript{2}) \\
Pyrgeometer Up (W/m2)   & Temperature 40 cm (°C)          & Snow depth (cm)                                        \\
Pyrgeometer Down (W/m2) & Temperature 80 cm (°C)          &                                                        \\
                        & Temperature 100 cm (°C)         &                                                        
\end{tblr}
}
\end{table}

\section{Methods}
\label{sec: methods}
\subsection{Physical basis}
\label{sec: fundamentos}
The insulation effect enabling temperature logger array methods to operate is closely related to the thermal diffusivity of snow, a parameter that varies with snowpack properties (such as density, water content, compaction or micro-structure). Typically, in numerical modelling, thermal diffusivity of snow is calculated in terms of various empirical parameterisations of thermal conductivity, which generally depend only on the density of the snowpack ($\rho$) \citep{SURFEX_diffusiv,diffusivity_neumann}.

Heat transport within the snowpack is primarily dominated by conduction, both through the solid ice matrix and through the air-filled pores. Additionally, latent heat exchanges and radiative heating also contribute to the overall heat flux. Other complex processes such as convection and wind pumping may also be important in certain situations \citep{diffusivity_neumann}. The dominant mechanism is then conduction through the ice matrix, as the thermal conductivity of ice is about two orders of magnitude higher than that of air. However, in warm snowpacks, latent heat transport can also play an important role. Both of these effects can be approximated through a heat diffusion equation that considers an effective thermal diffusivity coefficient since both possess a diffusive nature \citep{diffusivity_neumann}. The following heat equation can be written if only the vertical heat transport is considered:
\begin{equation}
    \frac{\partial T}{\partial t}=\alpha \frac{\partial^2 T}{\partial z^2}
    \label{eq: heat}
\end{equation}
where $\alpha$ [$m^2/s$] is the thermal diffusivity, $T$ is temperature, $t$ is time and $z$ is the vertical coordinate. Thermal diffusivity can also be expressed in terms of thermal conductivity ($\kappa$), density ($\rho$) and specific heat capacity ($c_p$): $\alpha = \tfrac{\kappa}{\rho c_p}$. 

Equation (\ref{eq: heat}) can be solved analytically if we consider harmonic boundary conditions or, in other words, considering the external (air) temperature signal is sinusoidal. Considering just one harmonic component (e.g. the daily cycle), a solution to the heat equation would be \citep{heat_eq_solution}:
\begin{equation}
T(z, t)=T_0+A_0 e^{-z \sqrt{\pi f_a / \alpha}} \cos \left(2 \pi f_a t-z \sqrt{\pi f_a / \alpha} + C_0\right)
\label{eq: sol_1harm}
\end{equation}
where $T(z,t)$ is the snow temperature at depth $z$ and time $t$, $T_0$ is the mean temperature, $C_0$ is a phase constant and $A_0$ and $f_a$ are the amplitude and frequency of the considered oscillation, respectively. From equation (\ref{eq: sol_1harm}) its clear that the initial oscillation undergoes a phase shift (term $-z \sqrt{\pi f_a / \alpha}$ inside the cosine) and an amplitude damping (term $e^{-z \sqrt{\pi f_a / \alpha}}$) with depth. This amplitude damping constitutes the principal effect that allows the identification of snow-covered, consequently facilitating the estimation of snow depth using a vertical array of temperature loggers. It is also clear that the amplitude attenuation and phase shift depend on the thermal diffusivity value $\alpha$ and thus on snow parameters such as density or water content.

A  limitation of solution (\ref{eq: sol_1harm}) is that temperature variation often differs from a simple harmonic oscillation. However, this can be addressed through Fourier analysis and by taking into account the linear properties of the heat equation, which enable a linear combination of (\ref{eq: sol_1harm}) to also be a solution for (\ref{eq: heat}). This solution in terms of a sum of Fourier harmonics can be expressed as follows \citep{heat_eq_solution}:
\begin{equation}
T(z, t)=T_0+\sum_{i=1}^N A_0 e^{-z \sqrt{\pi f_i / \alpha}} \cos \left(2 \pi f_i-z \sqrt{\pi f_i / \alpha}+C_i\right)
\label{eq: sol_Nharm}
\end{equation}
from where it can be easily seen that amplitude attenuation and phase shift increase with the frequency $f_i$ of each Fourier component. This indicates that the snowpack functions as a low-pass filter, dampening higher frequency variations to a greater extent.   

\subsection{Snow depth estimation}
\label{sec: methods_snow}

Snow depth estimates can be derived by analysing the temperature data at different heights above the ground recorded by the snow pole sensor. Fundamentally, the main idea is to identify which sensors are covered by snow and which ones are not. After this has been accomplished, the snow depth is assigned a value between the height of the uppermost covered sensor and the lowermost uncovered sensor. At the moment there are not, to our knowledge, methods that can determine precisely where the snow would lay between sensors. Other than interpolation or other more complex functions, a decision is usually made on where between sensors the estimated snow depth value will be assigned. These are usually the mid point between the uppermost covered sensor and the lowermost uncovered one or the height value of the uppermost covered \citep[e.g.][]{Lewkowicz2008}. We will use the former (hereafter referred to as the mid approach) and a different one (referred to as the upper approach) that considers the highest estimation, i.e. the height of the uncovered sensor closest to the snow. In section \ref{sec: results}, we will explain why we made this choice instead of considering the usual lower and mid values.

Various methods have been explored in order to determine which sensors are covered by snow and which are not \citep{danby,Lewkowicz2008,Lewkowicz_corr,alemanes,paper_luz}. These methods are largely based on the same basic idea: the reduction in snow temperature changes observed in measurements when a sensor is covered by snow versus when it is not. The quantification of this behaviour change can be made using various techniques. Most of them use the sub-daily temperature measurements at each height to calculate parameters such as daily thermal amplitude, variance \citep{danby,alemanes} or correlation \citep{Lewkowicz_corr} and then compare this values to some reference sensor assumed to be always uncovered or to some fixed value based on past records (climatology). Depending on how the series of the new parameter compares to the reference and given a certain threshold (hereafter referred to as the tolerance parameter) the sensors are classified as covered or uncovered. Different parameters can be chosen to quantify the thermal variations that a certain sensor records. Our analysis will demonstrate that the selection of a parameter and its time window can have a significant impact on the calculated snow depth. 

We will focus on comparing different choices of thermal variability quantifier parameters and their impact on the snow depth results. Each time a daily measure of the chosen parameter ($P_s$) is calculated from the temperature series. Then its value at every point of time is compared to that of the uppermost sensor in the vertical array setting (in the Crater Lake snow pole the sensor at height 160 cm, when it is functional). A tolerance parameter ($\gamma$) is chosen so that, when comparing with the uppermost sensor (reference), if a value verifies $P_s<\gamma P_{ref}$ then the sensor is considered covered. Note that other ways to determine if a sensor is covered have been proposed such as evaluating the difference of a chosen thermal variability statistic between adjacent sensors in the vertical and considering that the snow level lies between the sensors that show a greater difference. This is the approach proposed by \citep{alemanes} but it will not be replicated in this work. 

The different techniques used to quantify temperature variability in this study are detailed below:

\begin{itemize}
    \item \textbf{Method 1 (Daily thermal amplitude criterion)}: This first method is based on a basic quantifier of daily thermal variation, thermal amplitude. It is computed for each day and each series as $\Delta T = \max{(T)}-\min{(T)}$. As pointed out by \citep{Lewkowicz2008}, daily thermal amplitude is reduced if a sensor is covered by snow. This is a result of the amplitude damping of the thermal signal produced by the snow cover. Once the thermal amplitude of each sensor is computed it is compared to that of the uppermost sensor so that if $\Delta T_s<\gamma \Delta T_{ref}$ the sensor is labeled as covered. Daily thermal amplitude is one of the simplest metrics influenced by snow's thermal insulation, yet it disregards a significant amount of data that could prove useful, as it is based solely on the daily maximum and minimum temperatures.

    \item \textbf{Method 2 (Standard deviation criterion)}: Another way of characterizing thermal variability is through the standard deviation. A larger standard deviation indicates that the values present a broader spread around the mean of the data sample. When a sensor is covered by snow, its variation will be reduced as fluctuations are damped down by the thermal insulation effect, thus the standard deviation will be reduced. In \citep{alemanes} a 2-day variance is computed for each sensor. Then, inter-level differences are computed between adjacent sensors and the snow height is determined by the greatest drop in variance. Instead of variance, here its square root i.e. standard deviation is used. To ensure consistency with the other methods, instead of computing inter-level differences the comparison will be made by contrasting each series of standard deviation with the uppermost sensor. When those drop beyond a chosen tolerance value relative to that reference sensor the sensor is deemed to be covered. The advantage of using standard deviation over thermal amplitude is that it takes into account all data within the considered timeframe, also allowing for computation across different periods. Two possibilities are then considered: 

        \begin{itemize}
            \item \underline{Daily standard deviation}: the standard deviation is computed for each day considering only temperature measurements of that day. 
            \item \underline{Moving window standard deviation}: the standard deviation is computed for each day considering all values within centered multiple-day windows. This method facilitates the inclusion of more data, subsequently yielding less noisy results. On the other hand, this approach can prove counterproductive when the size of the window is excessively large, since it may fail to detect rapid fluctuations in snow depth. 
        \end{itemize}
    
    \item \textbf{Method 3 (Correlation criterion)}: This third criterion is based on \citep{Lewkowicz_corr}. Here, the criteria to determine if a sensor is covered is based on the correlation between time series taken by each temperature logger with respect to that taken by the uppermost sensor in the vertical array configuration. The correlation is computed via the squared Pearson correlation coefficient $r^2$ for each day in the time series, considering centered 11-day running windows. The correlation threshold was set empirically at 0.8, as specified in \citep{Lewkowicz_corr}, meaning that a sensor is considered covered by snow when its series $r^2$ with respect to the uppermost sensor series falls below 0.8.

    \item \textbf{Method 4 (Absolute rate of change criterion)}: This fourth criterion has been developed here and is based in the computation of a parameter that accounts for the rate of change per hour of the temperature data. The main idea behind this comes from the fact that Method 1 takes into account differences between just two of the daily data values, thus neglecting lots of other information from different data points other than the maximum and minimum. In order to solve this while still considering a quantifier based on temporal differences, instead of statistical parameters (like Methods 2 and 3) differences between adjacent in time data points are computed and divided by the time between observations (in our case $\Delta t =3$ hours). To benefit from the stability and robustness against noise that moving window methods offer, this differences are calculated in absolute value so that they can then be summed over the chosen window length. This results in a parameter like:
    \begin{equation}
    P =\frac{1}{N_{window}} \sum_{window} \frac{\left|T_i-T_{i-1}\right|}
    {\Delta t}
    \label{eq: param4}
    \end{equation}
    where $i$ corresponds to the time index. As it is written, equation (\ref{eq: param4}) corresponds to a moving average of $\frac{\left|T_i-T_{i-1}\right|} {\Delta t}$ which could be considered a numerical approximation (backward finite differences) of $\left|\frac{dT}
    {dt}\right|$. The parameter can then be interpreted as the moving window average of the absolute rate of change of temperature, measured in ºC per hour. As with other methods, once this parameter is computed for each series, sensors are considered covered if they verify: $P_s<\gamma P_{ref}$. 
\end{itemize}

\subsection{Spectral analysis and density estimation}
\label{sec: spectral_method}
As shown in section \ref{sec: fundamentos}, the amplitude damping and phase shift effects that the snow cover produces on the temperature signal detected by covered sensors are related via equation (\ref{eq: sol_Nharm}) to the signal frequencies ($f_i$), the snow thickness above the sensor ($z$) and the thermal diffusivity coefficient ($\alpha$). Given this relation, one could possibly use the temperature series at different heights recorded by the snow pole sensors to try to obtain an estimation of thermal diffusivity of the snowpack. Following this, one can invert one of the commonly used parameterizations of thermal diffusivity in terms of density to obtain approximate density values \citep{SURFEX_diffusiv,diffusivity_neumann}. 

Extracting a single frequency to be able to use equation (\ref{eq: sol_1harm}) is a difficult task in this type of data. Moreover, the most convenient one, the daily cycle, is not intense in half of the Antarctic year, so an approach that uses information from multiple frequencies is preferred. Such an approach was presented for soil temperature data in \citep{felix}. Given equation (\ref{eq: sol_Nharm}) and denoting as $P$ the periodogram calculated via Welch's modified periodogram \citep{welch}, spectral density attenuation with respect to a reference signal can be expressed as a function of depth and density as:  
\begin{equation}
\label{eq: power attenuation}
\zeta=\sqrt{P_z(f) / P_{ref}(f)}=e^{-z \sqrt{\pi f / \alpha}}
\end{equation}
where $f$ corresponds to the frequency of the different spectral components of the temperature signal $T(z,t)$. Taking the natural logarithm and squaring the equation, next relation is obtained:
\begin{equation}
(\ln{\zeta})^2= \frac{z^2\pi}{\alpha} f
\end{equation}
from which the value of $\alpha$ can be retrieved via linear fitting. In this case, the value of $z$ would correspond to the snow thickness between the sensors used for the computation of $\zeta$.

To guarantee values used to compute the linear fit are not excessively affected by noise some criteria are considered. Firstly, any spectral harmonics that have been attuned by more than $e^2$ times are removed, considering a thermal diffusivity value of $\alpha = 2.5\times10^{-7} m^2/s$ as given by \cite{diffusivity_neumann}. Subsequently, frequencies with an spectral power lower than a given threshold are removed from the computation, in order to suppress the possible errors of computing the ratio between two substantially low numbers. Subsequently, thermal diffusivity values are discarded if the Pearson correlation coefficient derived from the linear fit is not significant according to a t-test at a 99\% confidence level. 

Once values of thermal diffusivity are obtained they can be approximately converted to density via some of the usual parameterization used in snow models \citep{diffusivity_neumann,SURFEX_diffusiv}. The parameterization used has to be invertible for this to be possible. One of such parameterizations for thermal conductivity is the one from Yen (1981) \citep{yen1981review}:
\begin{equation}
k_{\text {snow }}=k_{\text {ice }}\left(\frac{\rho}{\rho_{\mathrm{w}}}\right)^{1.88}
\end{equation}
where $k_{\text {ice }}$ is the thermal conductivity of ice (2.22 W $\text{m}^{-1}$ $\text{K}^{-1}$), $\rho_w$ is the density of liquid water (1000 kg $\text{m}^{-3}$) and $\rho$ is the snow density. $k_{\text {snow }}$ is the thermal conductivity of snow that is related to its thermal diffusivity by its definition $k=\alpha\rho C_p$ where $C_p = 2090 \text{J kg}^{-1} \text{K}^{-1}$ is the specific heat of ice \citep{diffusivity_neumann}.

\section{Results and discussion}
\label{sec: results}

\subsection{Snow and temperature data}

\begin{figure*}[h!]
	\centering
	\includegraphics[width=1\textwidth]{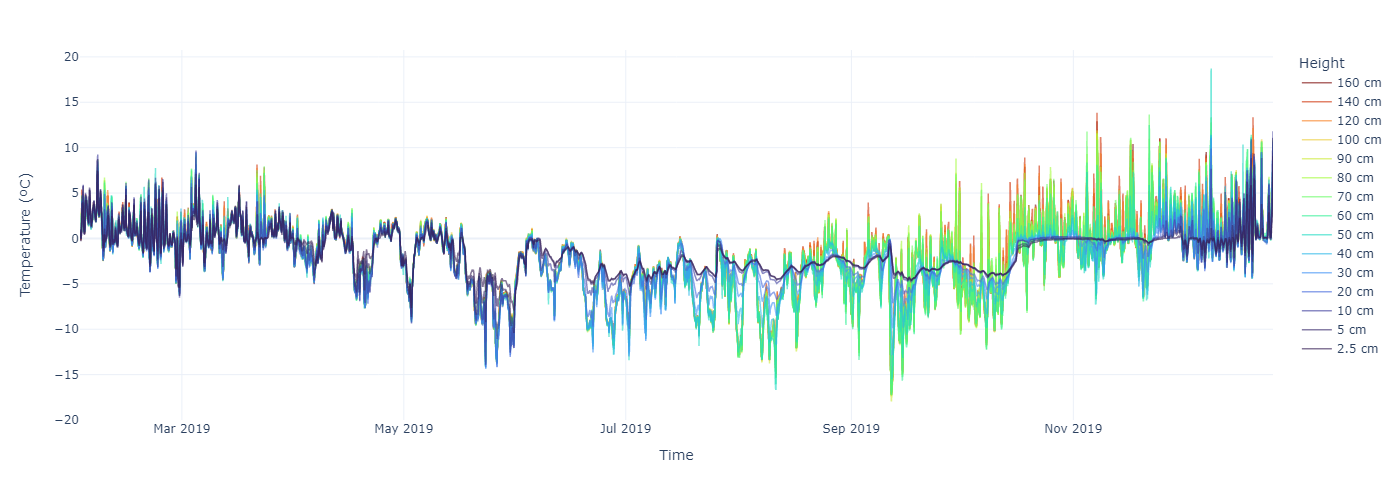}
	\caption{Time series of temperature at different heights above the ground recorded by the 15 sensors of the snow pole at Crater Lake during the year 2019.}
	\label{fig: temp_all}
\end{figure*}

Based on temperature data from the snow pole sensors (Fig. \ref{fig: temp_all}) and the SPA air temperature series for the years 2017 to 2023 (not shown), it can be seen that the temperature at the Crater Lake site has a clear annual cycle, with mean temperatures remaining below 0 ºC throughout the months of May to September-October. During the austral winter (months of June, July and August), even maximum temperatures do not rise above 0ºC. However, it was uncommon for air temperatures to fall below -15º C or exceed 10º C on Deception Island during the years that the SPA station was active.

\begin{figure*}[h]
	\centering
	\includegraphics[width=1\textwidth]{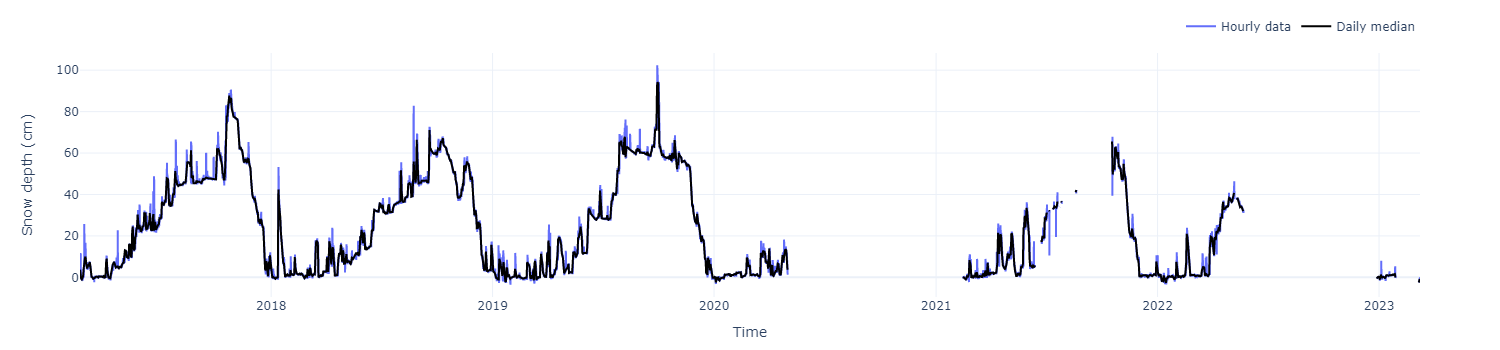}
	\caption{Raw hourly (blue line) and mean daily (black line) time series of snow depth recorded by the ultrasonic device of the SPA station between 2017 and 2023.}
	\label{fig: snow_spa}
\end{figure*}

The ultrasonic device at the SPA station has been gathering snow depth readings since 2017 (Fig. \ref{fig: snow_spa}), although complete snow depth series are only available for 2017, 2018 and 2019 due to the problems discussed in section \ref{sec: data}. The Crater Lake site has a seasonal snowpack, as shown in \ref{fig: snow_spa} and \cite{dePablo2020}, with an onset that tends to occur in late May to early June and an offset that typically occurs in mid-December. Generally, snow depth remains below 1 meter. 

\begin{figure*}[h]
     \centering
     \begin{subfigure}[b]{0.49\textwidth}
         \centering
         \includegraphics[width=\textwidth]{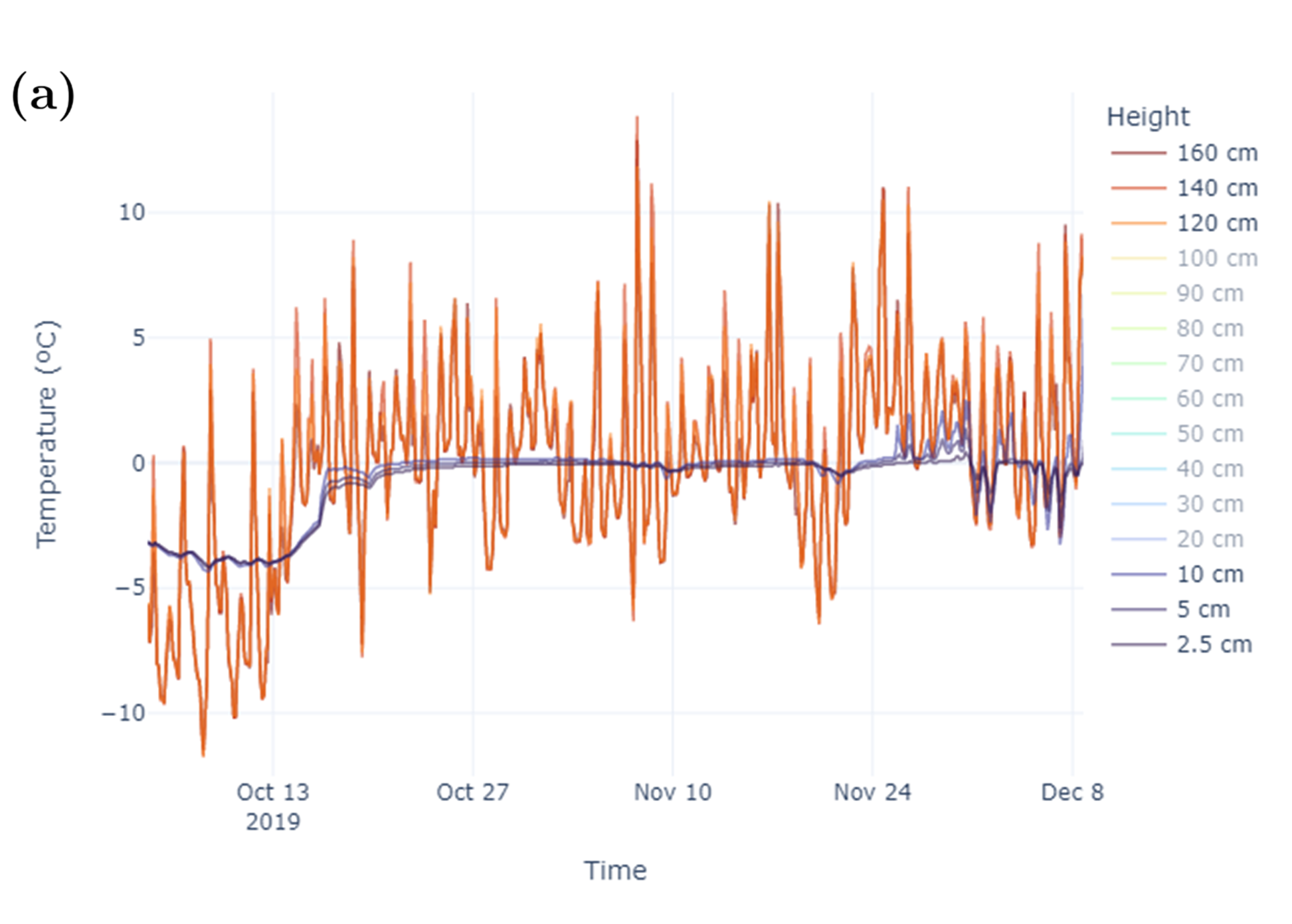}
         \label{fig: temp_curtain}
     \end{subfigure}
     \hfill
     \begin{subfigure}[b]{0.49\textwidth}
         \centering
         \includegraphics[width=\textwidth]{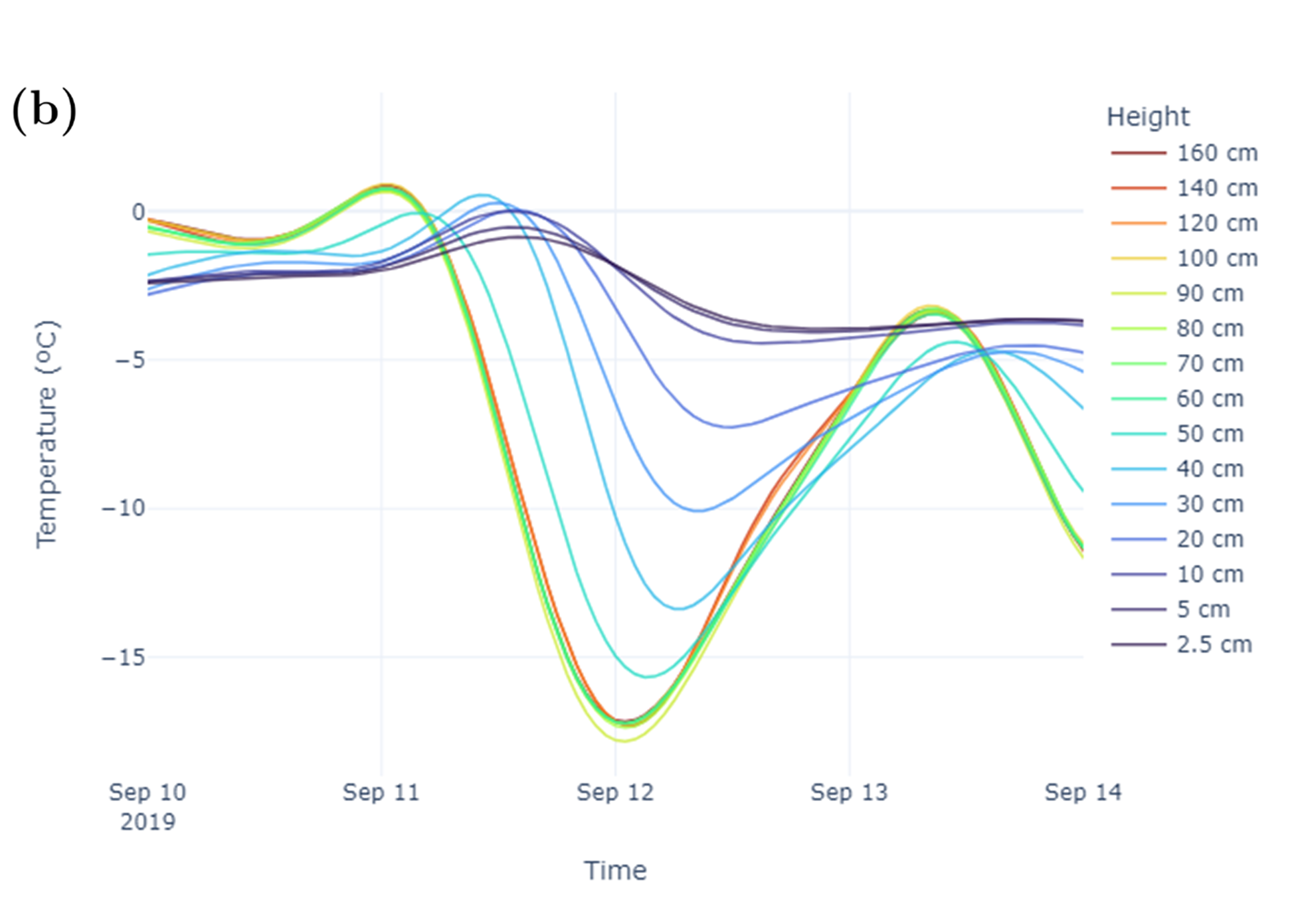}
         \label{fig: temp_desfase}
     \end{subfigure}
        \caption{Selection of some periods of the temperature data where relevant effects can be seen. (a): temperature series of the 3 uppermost and 3 lowermost sensors from October to early December where a zero-curtain effect can be seen on the lowermost sensors. (b): filtered temperature series for a 4 day period in September where an amplitude damping and phase shift between the uppermost sensors and the lowermost sensors can be identified. The data was filtered using a 22 hour cut low-pass Butterworth filter, to emphasize the daily cycle over sub-daily variability and noise.}
        \label{fig: temp_effects}
\end{figure*}

Temperature recordings taken at different heights provide insight into the impact of the snow cover (refer to Fig. \ref{fig: temp_all}). The effects of this cover are particularly evident during austral winter, when temperature records between different sensors exhibit clear differences. During that period sensors located closer to the ground (represented in colder colors in Fig. \ref{fig: temp_all}) show less thermal variation and less extreme temperatures than those situated at greater heights. This can be attributed to the insulation effect that the snowpack exerts in covered sensors. 

It is also apparent from Fig. \ref{fig: temp_all} that thermal amplitude exhibits a clear annual cycle even for sensors that remain exposed throughout the year (i.e. those located above 100 cm in Fig. \ref{fig: temp_all}). This cycle is due to the substantial difference in solar radiation reaching the site at different times of the year, with total daily values of downward long-wave radiation of less than 200 $W/m^2$ during austral winter compared to values of more than 10 times those during austral summer. This is due to the vast seasonal differences in the amount of daylight hours that the site's location presents, from less than 5 hours around winter solstice to more than 20 around summer solstice\footnote{Values taken from the website \url{https://www.timeanddate.com/sun/@6631955}.}. The reduced intensity of the daylight cycle could potentially affect the accuracy of snow depth estimation and especially hinder efforts to explore other applications of the snow pole, which may demand a more in-depth evaluation of temperature signals, as sometimes little to no daily cycle signal is present.

A relevant snow thermal behaviour occurs during the melting season (Fig. \ref{fig: temp_effects}a), when in mid October the air temperatures (represented by the 3 uppermost sensors) start to rise above 0ºC, allowing the snowpack to melt. Within days, the lowermost sensors' temperature also increases to nearly 0ºC and remains at a fairly constant value for more than a month. This phenomenon is commonly referred to as the zero-curtain effect \citep{zero_curtain2,zero_curtain}. At the start of the melting period, water starts to percolate through the snowpack and a layer of ice can form in the lowermost parts of the snowpack \citep{dePablo2020}. During this period, the snowpack temperature remains constant at around 0 oC as the heat uptake is used to melt the ice and not to raise the temperature. This means that latent heat exchange is dominant during this period and the snowpack heat transport cannot be assumed to be predominantly conductive.

The temperature data also show the amplitude attenuation and phase shifting effects of snow. This is particularly evident during periods of rapid air temperature change, such as the drop of more than 18ºC in 24 hours that occurred during 11-12 September 2019 (Fig. \ref{fig: temp_effects}b). Even with this important drop, it can be seen that temperatures stay a lot more stable at lower heights above the ground with a drop of just about 3ºC recorded by the  sensor at 5 cm height. The phase shift can also be seen, as after the minimum value was registered in the uncovered sensors almost 12 hours had to pass for a minimum to be reached in the sensor at 20 cm. All of this is due to the direct insulating effect of the nearly 60 cm snow cover that was present during that time.

\subsection{Snow depth estimation}
\begin{figure*}[h!]
     \centering
     \begin{subfigure}[b]{0.495\textwidth}
         \centering
         \includegraphics[width=\textwidth]{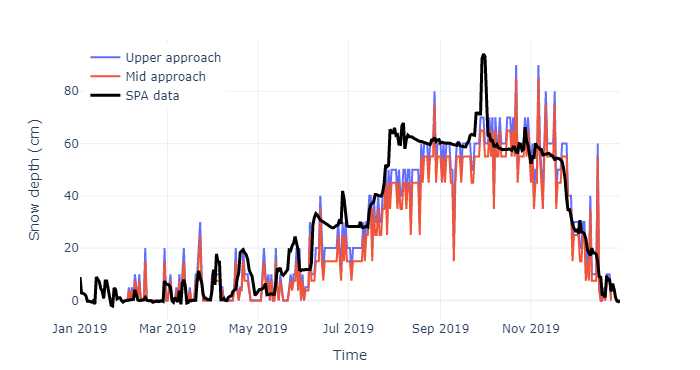}
         \caption{Method 1}
         \label{fig: m1}
     \end{subfigure}
     \hfill
     \begin{subfigure}[b]{0.495\textwidth}
         \centering
         \includegraphics[width=\textwidth]{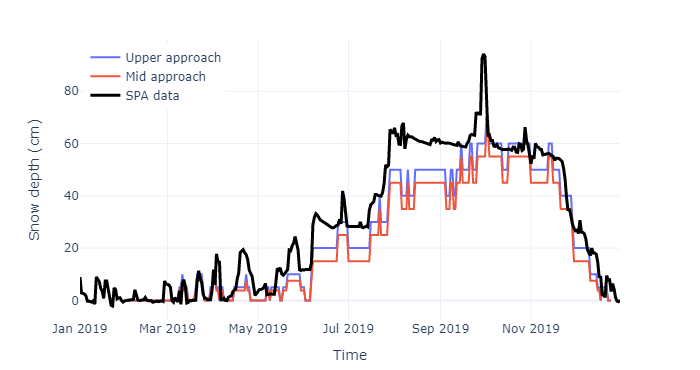}
         \caption{Method 2}
         \label{fig: m2}
     \end{subfigure}
          \hfill
     \begin{subfigure}[b]{0.495\textwidth}
         \centering
         \includegraphics[width=\textwidth]{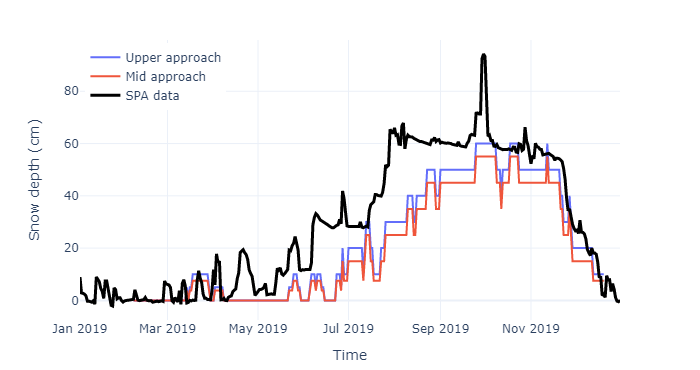}
         \caption{Method 3}
         \label{fig: m3}
     \end{subfigure}
          \hfill
     \begin{subfigure}[b]{0.495\textwidth}
         \centering
         \includegraphics[width=\textwidth]{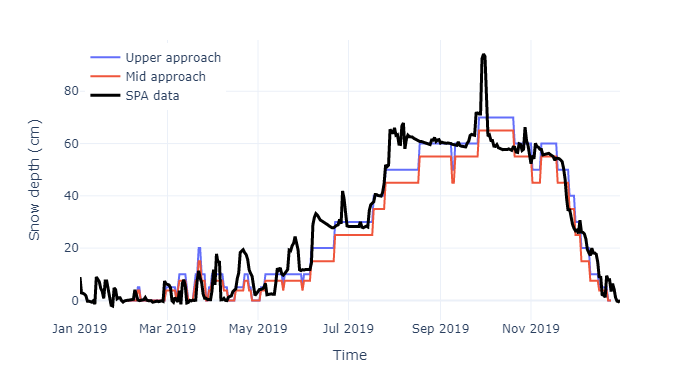}
         \caption{Method 4}
         \label{fig: m4}
     \end{subfigure}
        \caption{Comparison of snow depth obtained using 4 different methods (see details in section \ref{sec: methods_snow}) between data measured by the SPA ultrasonic sensor (black line) and estimated using two different approaches: considering the snow level halfway between the uppermost covered sensor and the next uncovered one (orange line), and considering the level of the lowermost uncovered sensor (purple line). The tolerance thresholds were chosen as 0.7 for the first two methods and 0.8 and 0.75 for the third and fourth, respectively.}
        \label{fig: all_methods}
\end{figure*}

Using the temperature series from the snow pole sensors daily snow depth values are estimated using the methods described in section \ref{sec: methods_snow}. For each of them, once the covered sensors are determined snow height values are obtained using both the upper and the mid point approaches (Fig. \ref{fig: all_methods}). 

\begin{figure*}[h]
	\centering
	\includegraphics[width=1\textwidth]{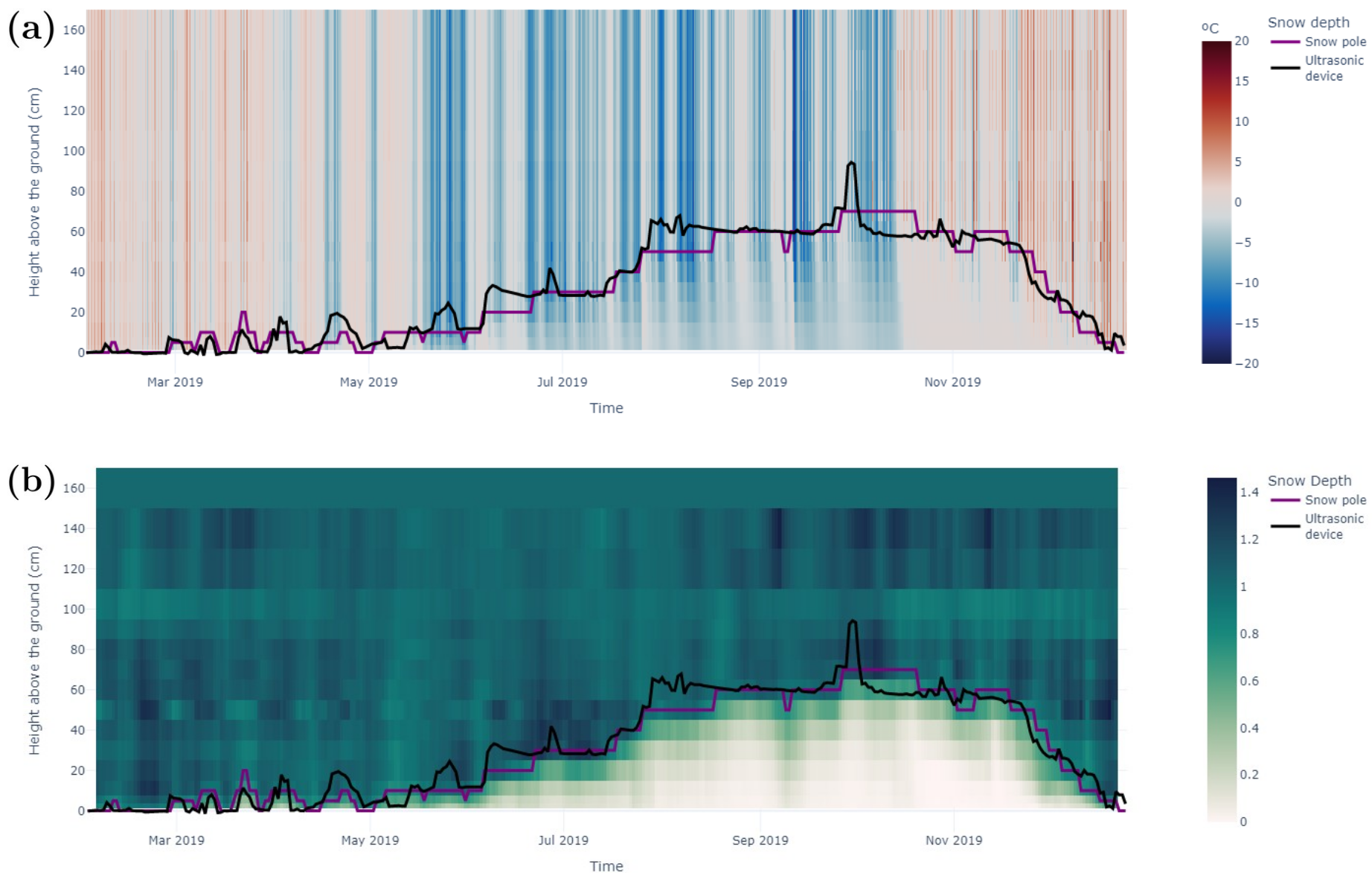}
	\caption{Comparison between best estimate snow depth using the snow pole data (purple line) and reference values from the SPA station (black line) and (a) a thermogram showing temperature evolution at each elevation (Y axis) along the time (X axis), and (b) the thermal variability quantifier used to determine if a sensor is covered by snow in Method 4 as described in section \ref{sec: methods_snow} relative to the value from the uppermost sensor. }
	\label{fig: m4_colors}
\end{figure*}

The first thing that can be noticed is that the difference between the two approaches (upper and mid) is just a shift of half the separation between adjacent sensors. This is not equivalent to a uniform shift in height of the whole estimation, as sensors are not equally spaced in the vertical. It can be seen that the upper approach is generally closer to the reference SPA data for all four methods under test. So, contrary to the logical believe that the middle ground between sensors would be the better approximation when we cannot know where between two adjacent sensors the snow lays, it seems that the highest possible value is the one that better resembles the reference. Values of root mean squared error and different comparisons will be made in the next subsection to assess the difference between approaches.

We now focus on the differences between each method. For the case where thermal amplitude (Method 1) was used the obtained estimation can be identified as the noisiest of them all. The estimation of snow depth considering the upper approach does lie relatively close to the reference but it presents lots of sudden oscillations that are unrealistic. This is probably due to it being the only one of the methods under test that does not use windowed variability quantifiers. As it uses just the daily difference between maximum and minimum values it seems to be subject to greater spurious fluctuations. For the second case a 7-days centered moving window standard deviation was computed to determine the covered sensors. It can be seen in the plot that using moving windows results beneficial, as unrealistic fluctuations are greatly reduced. The main problem of this second estimation is that it systematically underestimates the snow depth when compared to the reference value. This, and even in a worse matter, is also the case for the method using 11-day moving correlations like \cite{Lewkowicz_corr}. For all of the snow growing season this third estimation continuously undervalues the snow depth by 10 to 20 cm.

The forth estimation seems to be the closest to the reference SPA data as it benefits from the stability of using a moving window parameter and does not have the same problems of underestimation as the other methods. We will focus some of the forthcoming analysis on this forth method selecting the upper approach as we consider it the best approximation (further details are provided in the subsequent subsection).

Nevertheless, the fourth method of estimation also presents certain issues. It seems to fail to capture big increments in snow depth such as those occurring around the start of June and August. Additionally, it does not accurately capture brief, sharp peaks like the one observed in late September (lasting only three days). 

Focusing on this best estimate (method 4 considering the upper approach) it is apparent that the thermal regime of the site displays clear differences outside and inside of the snowpack throughout the snow season (Fig. \ref{fig: m4_colors}a). Starting mid May the snowpack begins to grow as temperatures go below the freezing point, allowing snow to fall and settle. During this period, the effects of thermal insulation are noticeable, as the temperature variability inside the snowpack decreases (amplitude damping), and signals from strong air temperature variations shift to the right (phase shift). A change in the thermal regime can be identified around mid-October as the air temperatures start to reach above 0°C again. Inside the snowpack, the variability is further reduced as the snow starts to melt, and liquid water percolates the snowpack. This results in most of the snowpack maintaining a constant temperature around the freezing point, which is known as the zero-curtain effect (Fig. \ref{fig: temp_effects}a) \citep{zero_curtain2,zero_curtain}. 

The examination of the selected parameter in relation to the uppermost sensor values aids in explaining the criterion for identifying covered sensors (Fig. \ref{fig: m4_colors}b).  When the value drops below the tolerance threshold, we consider the sensor to be covered. As a result, the snowpack surface displays a significant reduction in value, indicating that all sensors beneath it are covered.

\begin{figure}[h!]
    \centering 
    \includegraphics[width=0.45\textwidth]{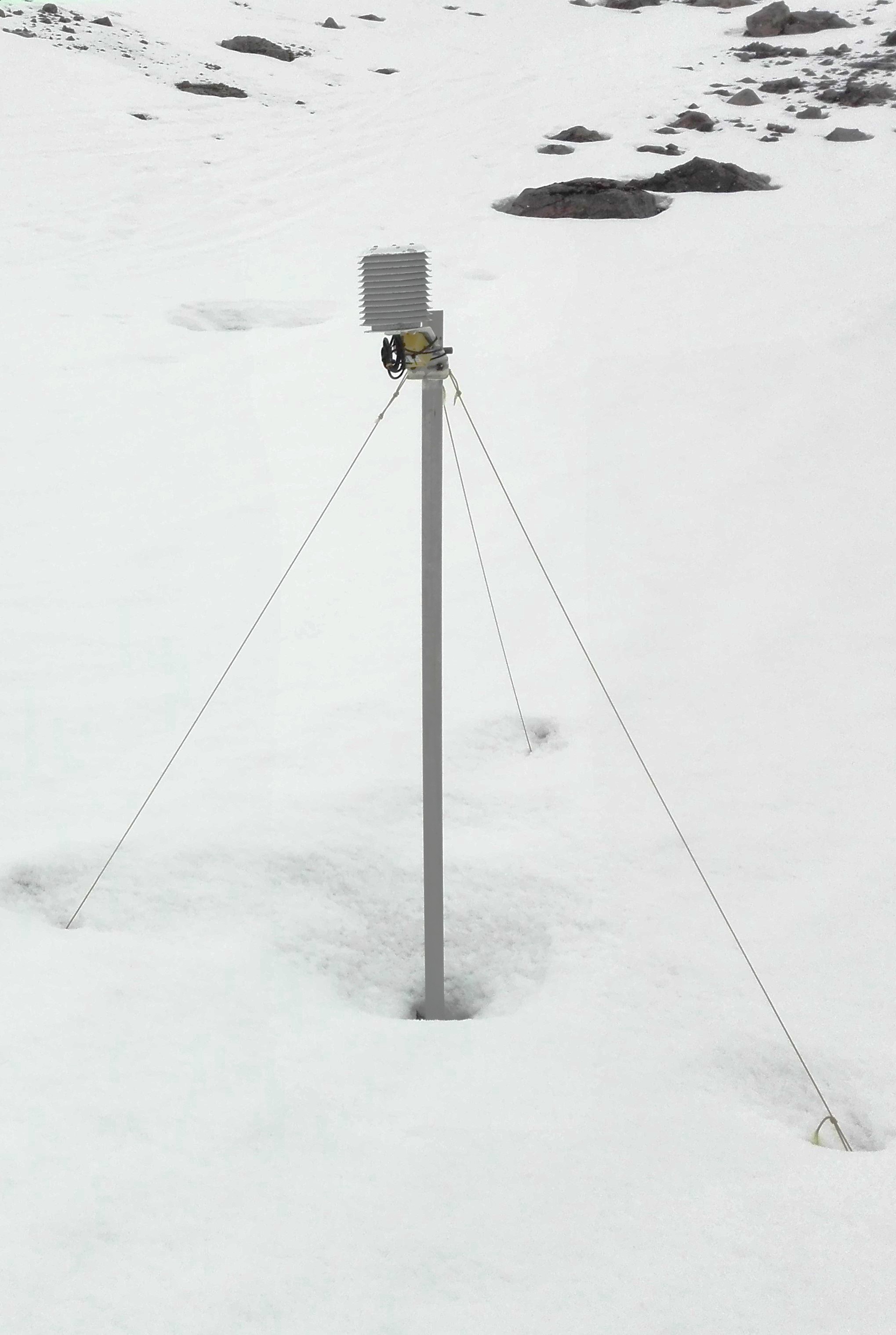} \caption{Picture of a pole supporting an air temperature logger installed at a different study site, depicting the formation of a ``pit'' of lower snow depth around its base due to long wave radiation from itself.}    
    \label{fig: pit} 
\end{figure}

Some of the discrepancies detected in the estimation may be ascribed to swift and sudden changes in temperature, as their magnitude can, at times, cause them to spread into the snowpack and cause false reductions in the assessed snow depth. For example the -18º C drop registered during mid September (Fig. \ref{fig: temp_effects}) alters the value of the thermal variability quantifier (Fig. \ref{fig: m4_colors}b), as the change in temperature is sufficient to increase the variability value of some of the sensors inside the snowpack. We can also clearly see that the main errors that are still present in this estimation (underestimation of the snow growth around the start of June and August) manifest here in the fact that the values of the chosen variability quantifier associated to those sensors that should be detected as covered is still as high as that of the uncovered sensors. We believe this could be due fresh snow being less dense and more conductive or to the effect depicted in Fig. \ref{fig: pit} where snow melts more around structures because of thermal radiation emitted by them generating a pit that would reduce the snow depth just around the snow pole but not further apart (where the reference ultrasonic device takes its measurements). Fig. \ref{fig: m4_colors}b also show some differences between sensors that are uncovered. For example sensors at 120 and 140 cm of height seem to detect changes in a more abrupt manner than the uppermost sensor and the sensor at 50 cm also seems to behave differently at some periods. Such discrepancies could be due to differences in calibration between the sensors or to differences in exposure to the sun, as some sensors may be obstructed by other elements of the SPA station collocated near them (see Fig. \ref{fig: SPA_photo}). It may be worthwhile to conduct a detailed analysis of these discrepancies to determine if they might negatively impact the outcomes and to explore potential measures for minimising their impact. 

\subsection{Error assessment}
\subsubsection{Methods and approaches}
In order to analyse the differences between the various methods and approaches used (see Fig. \ref{fig: all_methods}), the differences between the estimates and the reference data are compared mainly by calculating the root-mean squared error (RMSE) and the Pearson correlation coefficient. 

\begin{figure}[h!]
     \centering
     \begin{subfigure}[b]{0.485\textwidth}
         \centering
         \includegraphics[width=\textwidth]{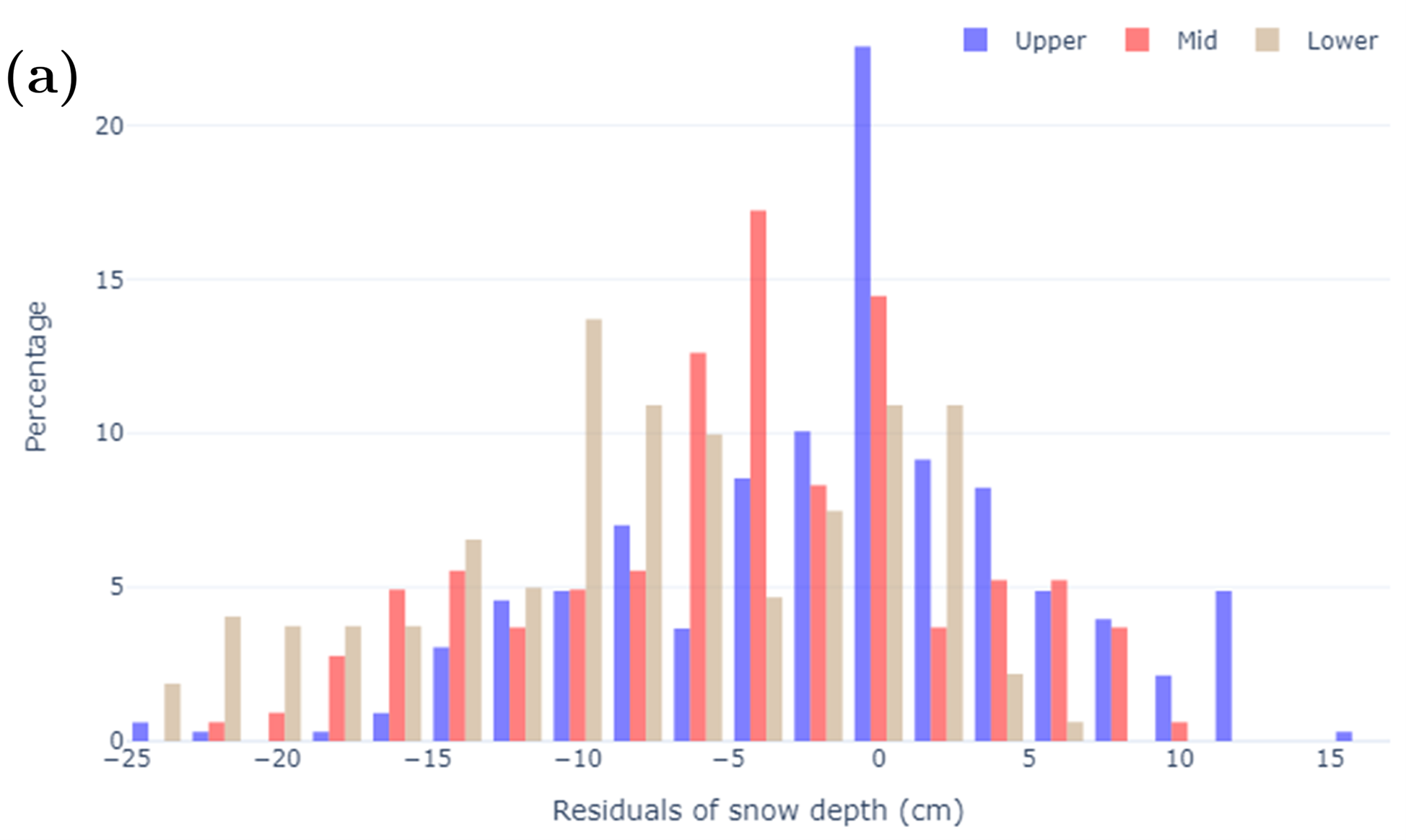}
         \label{fig: residuals}
     \end{subfigure}
     \hfill
     \begin{subfigure}[b]{0.485\textwidth}
         \centering
         \includegraphics[width=\textwidth]{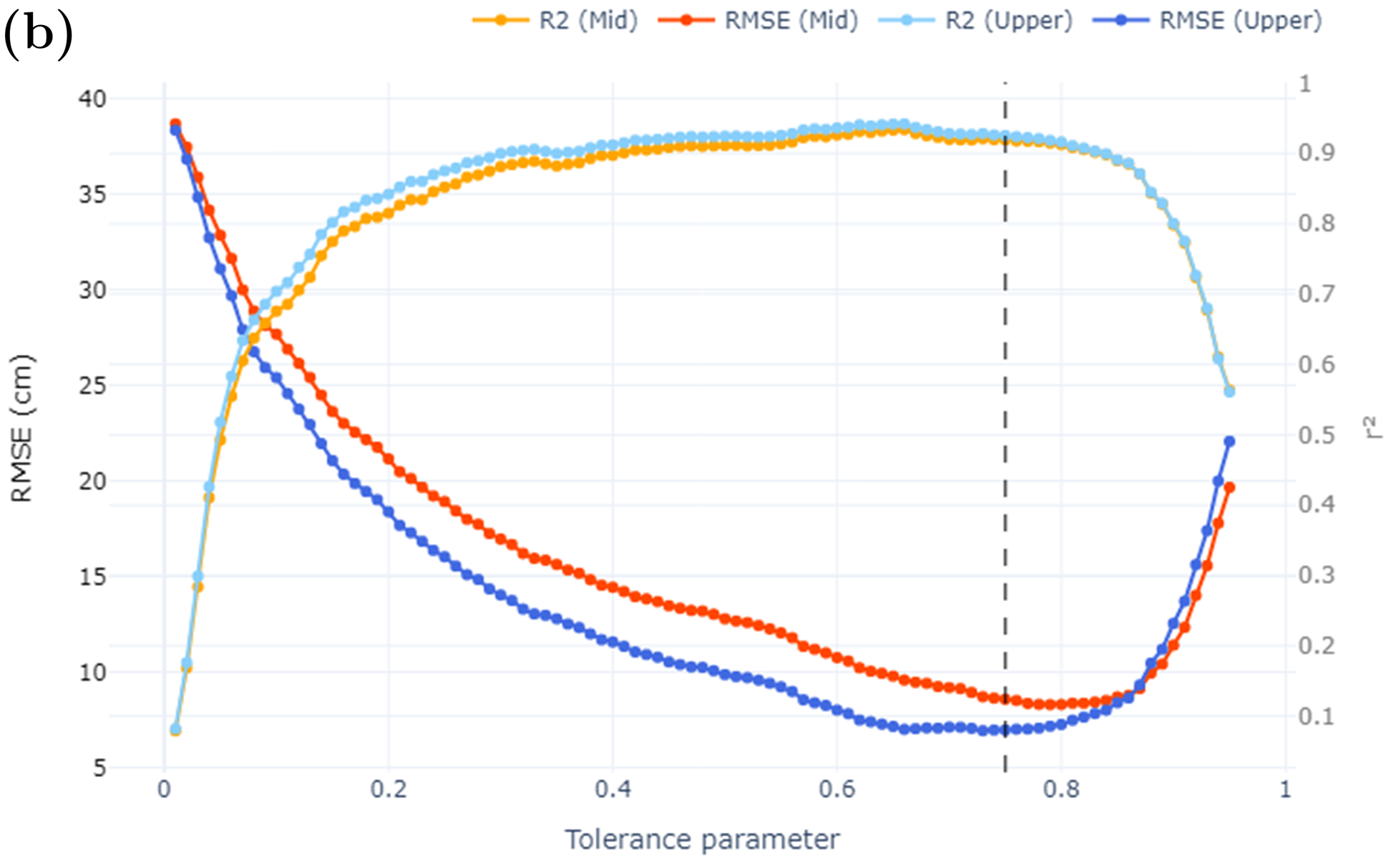}
         \label{fig: m4_erros}
     \end{subfigure}
        \caption{(a): residuals of Method 4 with 0.75 tolerance parameter snow depth approximation with respect to the SPA station reference data for the upper level (purple), the mid level (orange) and the lower level (brown) approaches. (b): comparison of root mean squared error (RMSE) and squared Pearson correlation coefficient of the upper and mid point approaches considering Method 4 for different values of the tolerance parameter used to decide whether sensors are covered or not. The value used for Fig. \ref{fig: m4} is indicated by a black dashed line.}
        \label{fig: m4_err_res}
\end{figure}

\begin{figure*}[h!]
	\centering
	\includegraphics[width=1\textwidth]{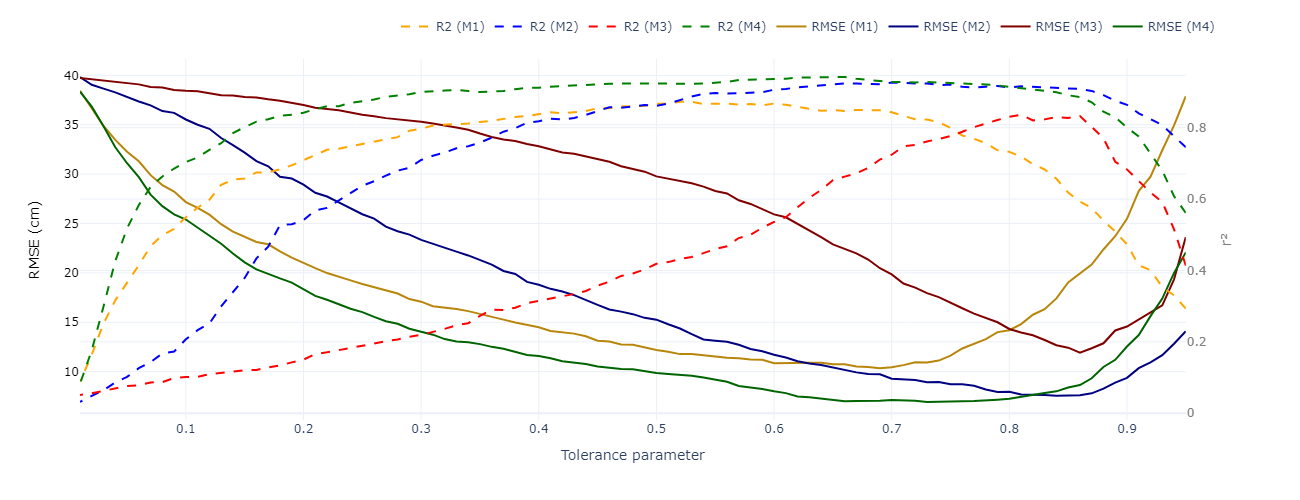}
	\caption{Root mean squared error (RMSE) and squared Pearson correlation coefficient of all four considered methods to obtain the snow depth estimation versus value of the tolerance parameter used in each method to determine if a sensor is covered compared to the uppermost sensor. In all cases the upper level approach is considered.}
	\label{fig: tol_all}
\end{figure*}

While Method 4 provided the closest estimation to the reference data (see Figure \ref{fig: all_methods}), the negative residuals observed in comparison to the reference data indicate that underestimation is still present. The distribution of the residuals varies greatly depending on the approach chosen when deciding where to determine the snow depth between sensors (\ref{fig: m4_err_res}a). The mid point approach systematically underestimates the snow depth as its residuals distribution peaks at negative values and has a mean of -4.96 cm. The lower point case shows an even worse underestimation having its mean at -8.15 cm. This is the main reason why it was excluded from previous figures, as its results always present clear bias. In contrast, the error distribution of the upper approach has its peak around zero and its mean, although still at negative values, is much closer to zero at -1.23 cm. Both the mid point and the upper point cases seem to present a long-tailed distribution at negative values that can be associated to underestimation at certain periods (see Fig. \ref{fig: m4}). An association can be made between those underestimations and the growth rate of the snowpack. Calculating absolute height changes in the reference and averaging them over moving windows, along with computing RMSE in those same moving windows, reveals a significant\footnote{Significance of the correlation coefficient is determined by a two-sided t-test with at least 90\% confidence.} correlation between the two variables with a Pearson correlation coefficient of 0.5. This suggests that the algorithm is mainly struggling with rapid changes in snow depth during sudden growth events. 

A comparison of the RMSE and the squared Pearson correlation coefficient (Fig. \ref{fig: m4_err_res}b) shows that both values improve as the tolerance level increases, but start to deteriorate at around 0.8. The optimum correlation peak and the lowest RMSE occur at tolerance levels between 0.6 and 0.8. Therefore, the chosen value of 0.75 is justified for the estimation of the snow depth in Method 4 (Fig. \ref{fig: m4}). The upper approximation presents higher or equal values of correlation than the mid point one while it also presents a lower RMSE for most tolerance values. This supports the decision to employ the upper point approximation to evaluate further the performance of the estimation and the thermal regime of the snowpack (Fig. \ref{fig: m4_colors}). The minimum value for the RMSE corresponds to the upper approach with a tolerance of 0.75 and has a value of 6.97 cm. It is noteworthy that this figure is significantly lower than the average gap between sensors of 11.25 cm. 

The fact that the mean of both residuals distributions is still at negative values and that the upper approach gives better results than the more commonly used mid point approach (Fig. \ref{fig: m4_err_res}) could indicate that a small difference between the zero level of the reference ultrasonic device and the snow pole might exist since they are not measuring exactly at the same point (as seen in Fig. \ref{fig: SPA_photo}). To explore this hypothesis, we performed RMSE and Pearson correlation analyses introducing different offsets to the estimates (not shown). The addition of an offset did indeed improve the estimation of the upper approach, reducing its RMSE as the offset is increased until a value between 1-1.5 cm, but the reduction in RMSE is less than -0.1 cm. A greater offset worsens the upper approach but improves the middle approach until a value of 5-5.5 cm is reached, but this reduction is not sufficient for the mid point approach to give RMSE values less than 6.97 cm, so it does not explain why the mid point approach gives worse results than the upper approach.

Another hypothesis for the advantage of the upper approach could be related to the low spacing between the sensors in this snow pole compared to other more conventional designs. This may imply that a minimum quantity of snow is required above a sensor to register it as covered. Therefore, when sensors are positioned closely together, underestimation may occur but can be compensated for by skipping to the next sensor. In cases where the vertical resolution is coarse, this skip may be too large and the mid-point approach may be preferred.
Further research using alternative equipment is necessary to establish the correlation between these findings and the vertical configuration of the snow pole.

A comparison of the RMSE and Pearson correlation between four methods was conducted using the upper approach (Fig. \ref{fig: tol_all}).  It was found that Method 4 displayed a higher correlation coefficient with the reference for most tolerance values. Conversely, Method 3 demonstrated relatively low correlation coefficients for most tolerance levels, thus emerging as the least reliable option. About the RMSE almost the same conclusions can be drawn, as the lowest RMSE for most tolerance values corresponds to Method 4 and the highest to Method 3. There are also clear differences between methods regarding the best value of the tolerance parameter that can be used, since Method 1 shows values around 0.7 while for the Method 3 is about 0.85. The main problem with Method 1 was that it presented spurious oscillations in snow depth that may result from daily thermal amplitude being more variable than moving window-based parameters. It makes sense that its best tolerance value would be the lowest among all methods, as increasing it might generate even more sensibility to spurious changes. Method 2 seems to be almost on pair with Method 4 when considering a higher tolerance than 0.8 but its lowest RMSE is still a bit higher than the best value from the Method 4.

\subsubsection{Missing sensor sensibility analysis}
\begin{figure}[h!]
	\centering
	\includegraphics[width=0.48\textwidth]{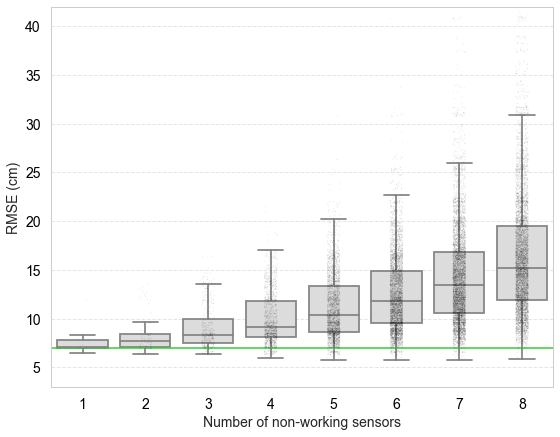}
	\caption{Boxplot of the RMSE obtained using Method 4 and the upper approach with a tolerance value of 0.75 if all possible combinations of a certain number of sensors are deleted from the 2019 database. Grey points represent individual values. The RMSE of the base case with all 15 sensors working is represented as a green line corresponding to 6.97 cm}
	\label{fig: boxplot}
\end{figure}

As it was commented in section \ref{sec: data} another advantage other than price the snow pole system possesses over ultrasonic devices would be that a malfunction of some of the sensors in a snow pole will still permit an estimation of the snow depth to be obtained, contrary to when the ultrasonic device or its batteries malfunction and complete loss of data is possible. Nevertheless, it is expected that a malfunction of some of the sensors in the pole will worsen the snow depth estimation. For instance, applying the same considerations as for the best estimate of snow depth in 2019, which gave a RMSE of 6.97 cm, but for the year 2018 (when sensors at 5, 10 and 40 cm did not work) gives a RMSE of 7.86 cm, less than 1 cm worse than the 2019 estimate. As the 2018 snow season may have had different characteristics that could make the comparison with 2019 inaccurate, some sensitivity tests were performed on the 2019 dataset. Data from certain sensors was omitted to test how the RMSE changes compared to the full estimation with all sensors working, considering the distribution of RMSE for all possible combinations of a given number of missing sensors (Fig. \ref{fig: boxplot}). As more sensors are eliminated, the median RMSE increases compared to 6.97 cm. The borderlines of each box, representing the 25th and 75th percentiles, also increase. Still, for the last 5 years of data only 1 to 3 sensors failed each year, so the median RMSE of estimations would not have worsen more than 2 cm. This outcome is preferable to a complete loss of snow depth data.

Although data loss from some sensors generally worsens results, there are points within the range of possible combinations that can improve RMSE. This may be due to certain sensors having problems that worsen results, but it appears that combinations of more separated sensors yield coarser estimations that may be more accurate in terms of RMSE, but not necessarily in terms of correlation or other error quantifiers. Further investigation is required to determine if there is a pattern or link to specific sensors in cases where the RMSE shows improvement.

\subsection{Spectral analysis and density estimation}
\subsubsection{Time series spectral analysis}
Spectral analysis was conducted on the temperature time series data from each sensor on the snow pole to investigate the alterations in spectrum caused by the snow cover and to evaluate certain features of the Antarctic environment. 

\begin{figure*}[h!]
\centering
\includegraphics[width=1\textwidth]{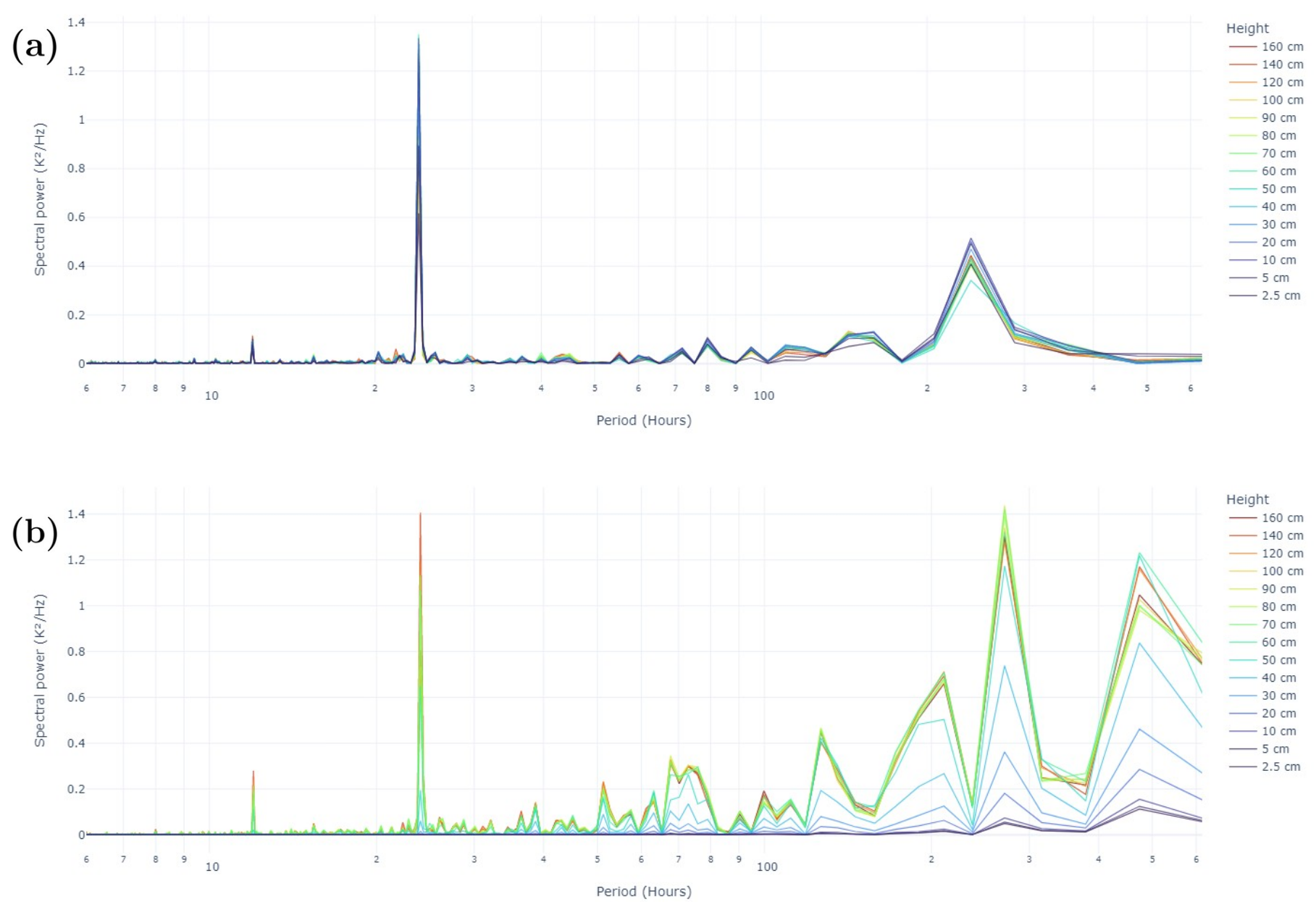}
\caption{Periodograms of the temperature signals at different heights from the snow pole sensors, with the X axis converted from signal frequency to period represented in logarithmic scale, for a selected period: (a) from 01-02-2019 to 01-04-2019, a period when no seasonal snow cover was present but when some smaller snowfall events did occur;  (b) from 29-07-2019 to 15-10-2019, a period when a consistent seasonal snow cover of more than 50 cm was present.}
\label{fig: spectrums}
\end{figure*}

\begin{figure*}[h!]
	\centering
	\includegraphics[width=1\textwidth]{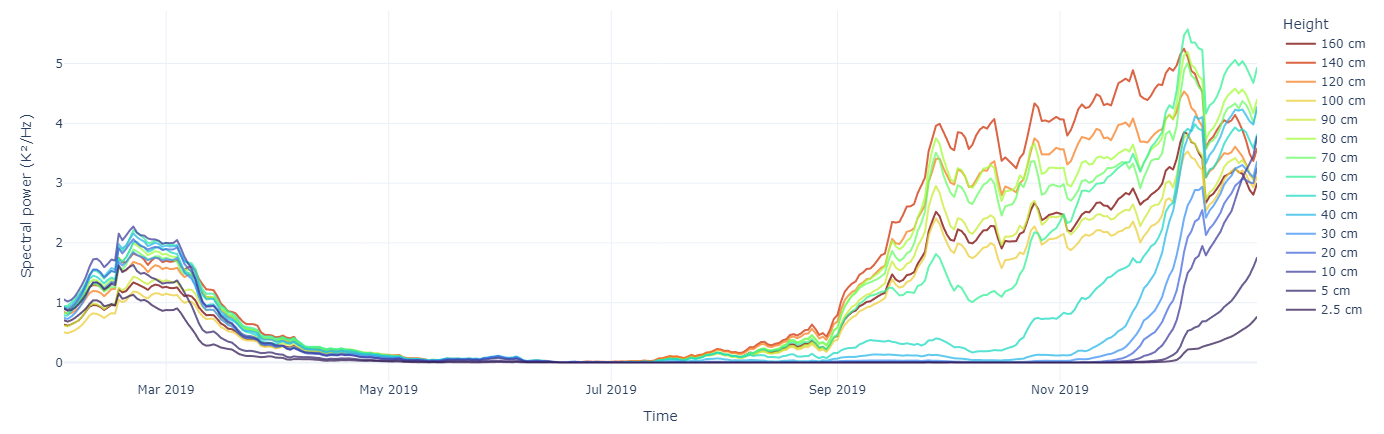}
	\caption{Daily time series of spectral power density associated with the 24 hours period (daily cycle). Values correspond to 30-day moving window calculation via the periodogram.}
	\label{fig: 24h}
\end{figure*}

\begin{figure*}[h!]
	\centering
	\includegraphics[width=1\textwidth]{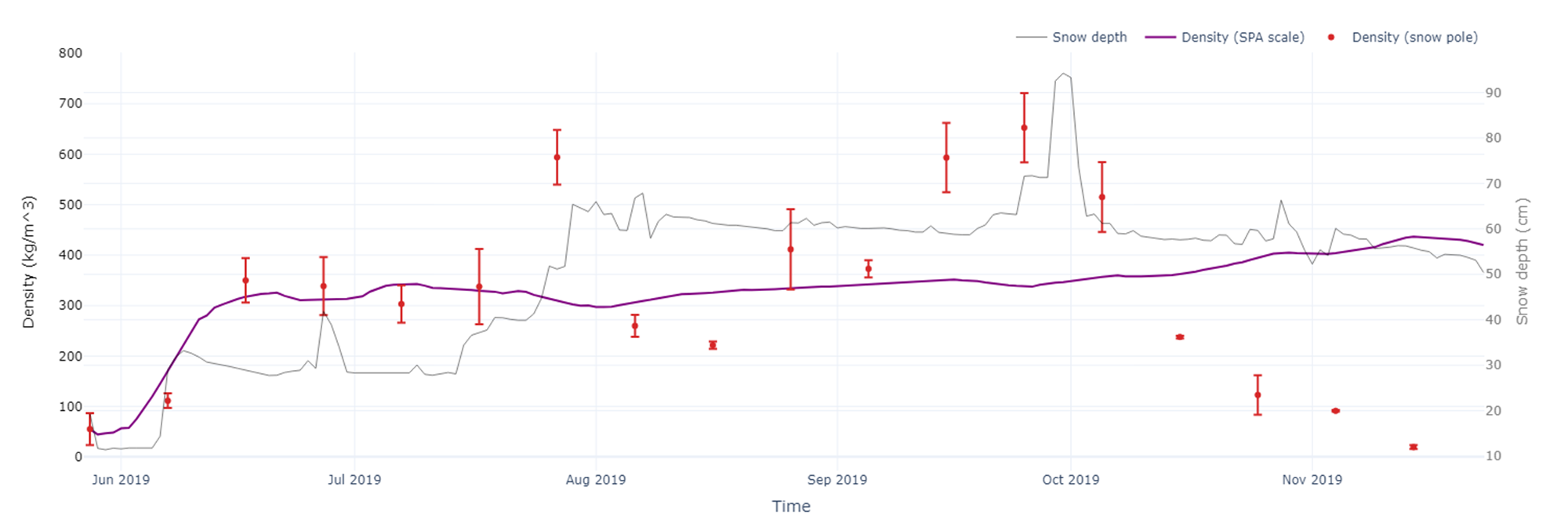}
	\caption{Approximate density values retrieved through a 20 day moving window spectral density attenuation analysis smoothed as 10-day averages (red points) with their standard error of the mean (whiskers), compared to density time series derived from the SPA station ultrasonic snow depth sensor and the snow scale (purple line). SPA snow depth values from the ultrasonic device (black line) is shown as reference of snow evolution.}
	\label{fig: density}
\end{figure*}

During the Antarctic summer and early autumn season, periodograms are similar because snow cover is not yet present or is discontinuous and disappears after a few hours (Fig. \ref{fig: spectrums}a). The dominant frequency is that associated with the diurnal cycle, and a secondary frequency is also present at periods of about 10 days, usually associated with synoptic variability. However, this pattern is not observed during the winter and early spring seasons (Fig. \ref{fig: spectrums}b). During this period, sensors located below 50 cm were usually covered due to the presence of the seasonal snowpack. As a result, their spectral power densities are attenuated with respect to the external signal. This attenuation is clearly more pronounced for higher frequencies (lower periods) as predicted by equation (\ref{eq: power attenuation}). This effect is especially evident for the lowermost sensor, whose power spectrum at high frequencies quickly becomes almost flat. These findings support the idea that the snow acts as a low pass filter, attenuating more at higher frequencies (Figure \ref{fig: spectrums}b).

In order to use this spectral analysis to assess one of the exceptional characteristics of the Antarctic environment, we analyse the temporal evolution of the power density associated with the diurnal cycle (Fig. \ref{fig: 24h}). The results indicate that the daily cycle is notably weak and almost absent for half of the year, particularly during the austral winter. This is an important limitation when using this cycle to study changes such as amplitude attenuation or phase shift to obtain information from the snowpack, as most of the snow season coincides with this low light period.

\subsubsection{Snow cover density calculation}

To examine the feasibility of obtaining acceptable density approximations through spectral analysis, we firstly approximated the density values for a whole period identical to that of the winter/spring period (Fig. \ref{fig: spectrums}b). Results were compared to a mean density for the period of 336.2 kg $\text{m}^{-3}$ calculated directly from the SPA snow depth and snow weight per square metre data. The best results were obtained when lower sensor pairs were used. For example, using the sensors at 5 and 10 cm provides a density of 388.9 (323.3, 485.2) kg $\text{m}^{-3}$ where values in parentheses correspond to 95\% confidence intervals, meanwhile the 2.5 and 5 cm pair gave a value of 429.6 (313.3, 667.0) kg $\text{m}^{-3}$ and the 2.5 and 10 cm one yielded 398.2 (355.4, 451.8) kg $\text{m}^{-3}$. The inclusion of sensors at higher heights in the considered pairs resulted in density values higher than expected, in many cases exceeding 1000 kg $\text{m}^{-3}$, which corresponds to the density of liquid water. These results are nonphysical and, in conjunction with the minimal linear fit coefficients obtained from these upper sensor pairs, indicate a flaw in the methodology. Further investigation is required to identify the causes of these issues and to develop possible solutions. Nonetheless, it is noteworthy that temperature sensors alone can provide a near-accurate estimation of snowpack density.

Secondly, density values were retrieved for 20-day windows using sensor pairs below 15 cm for the entire period when snow depth was greater than 15 cm. The results of the three possible pairs were combined into a mean weighted by both the Pearson correlation of the linear fit and the standard error of each derived value. As a higher-than-expected variability was observed in the resulting density series, results where pulled into centered 10-day window averages (Fig. \ref{fig: density}). The smoothed results still have a higher than expected variability when compared to the density series derived from the snow depth and snow weight. Further investigation is required into potential factors that may cause the estimation to be overly sensitive to changes. However, it should be noted, as an initial observation, that abnormal increases in the estimation are correlated with significant snowfall occurrences.

During the snow melt period starting mid October an abnormal drop is visible in the estimation (Fig. \ref{fig: density}). Such a drop may be attributed to the zero-curtain effect and the percolation of liquid water into the snowpack as it melts. During that period studies relaying on the thermal regime being conductive might be seriously impaired. Notably, between June and July, the estimation values are generally consistent with those obtained from depth and weight measurements. It is worth noting that the reference values relate to the entire snowpack whereas the estimated ones only utilise data from the lower 10 cm.  Nonetheless, despite the methodology being a mere adaptation of a technique utilised in soil studies \citep{heat_eq_solution,fourier_ml,felix}, these findings provide optimism that a refined approach accounting for snow-specific processes could achieve proficient density estimations utilising solely temperature data at various heights. 

\section{Conclusions}
This study derives snow cover thickness utilizing four distinct methods, each based on the thermal behaviour of air temperature measured at varying heights above the ground and the changes in snow depth due to the insulating effect of the snowpack. After classifying the sensors in the snow pole as either covered or uncovered, the depth of the snow cover was determined using three distinct approaches in each method: taking into account the height of the uppermost sensor covered by snow (Low), the lowermost uncovered (Upper), or the mean between them (Mid). These findings were then compared to high-resolution snow depth measurements obtained via an ultrasonic sensor installed at the same location. Calculations have been made to assess errors in various approaches, and additional analyses have been conducted to investigate other snow cover parameters, including snow thermal diffusivity and density. The conclusions are as follows:
\begin{itemize}
    
    \item Although all tested methods demonstrate a snow cover development that corresponds with the reference, the majority present issues such as underestimation. Even the most accurate estimate shows problems of underestimation during the growing season, especially in relation to the first days after heavy snowfall events and with increases lasting only a few days. 
    \item A new modified estimation method for snow depth was developed using the moving average absolute rate of change as a quantifier of thermal variability, which gives the lowest RMSE of all the methods tested and a high correlation with the reference values. 
    \item  Algorithms utilizing windowed variability quantifiers obtain a more accurate depiction of snow depth. Unnatural oscillations are observed in the estimated snow depth when non-windowed quantifiers are employed, requiring filters or moving averaging to be appropriate. 
    \item Unlike previous studies, which use the midpoint approach to assign snow depth once the covered sensors have been determined, the findings presented here demonstrate that the upper approach achieves better alignment with the reference snow depths. This may be related to the unusually high vertical resolution of the snow pole being examined, in contrast to other more frequently used installations.
    \item A snow pole with a set of temperature sensors has been shown to be reliable in providing an accurate estimate of snow depth compared to the ultrasonic sensor, and appropriate results could still be obtained even if individual sensors failed. Snow poles offer numerous advantages, including their economical, logistical, and methodological benefits, for monitoring snow depth in remote and extensive areas, compared to high technology and complex monitoring instruments.
    \item There are some remaining issues that require attention regarding the snow pole system. For example, the structure of the pole itself can affect the surface, leading to partial melting of the snow cover. Additionally, the presence of adjacent equipment or differences in sensor orientation can create discrepancies. Furthermore, anomalously high thermal variability events for the region can propagate through the mantle and cause errors in the estimates.
    \item Data collected from snow poles show the phase shift and spectral amplitude attenuation caused by snow cover in the temperature signal. This enables the development of techniques to calculate the density of snow coverage by using spectral amplitude attenuation. A method was tested, demonstrating preliminary yet promising results, although it significantly under-performs during the melting season. Further research is needed to establish a valid methodology, but this may unveil new potential applications for the snow pole beyond snow-depth estimation. 
\end{itemize}

Future work needs to be conducted on snow depth estimation errors and their specific causes. Furthermore, additional methods for retrieving density information should be explored. The utilization of phase shift could also be beneficial in determining new data, as most techniques only consider signal attenuation whilst discarding any phase shift. It is also necessary to investigate the usefulness of snow poles outside polar regions, as a higher variability in external temperature signals could potentially improve or worsen results and, in temperate latitudes, certain snow regions may be more affected by above-zero air temperatures. Insights from this study could be useful for the future construction of new snow pole equipment, which could extend the usefulness of these devices to other regions or to measure other snow properties. 

\section{Aknowledgements}
The authors thanks chiefs and crews of the Spanish Antarctic Station “Gabriel de Castilla”, in Deception Island, for their support during field works and help on the instrument’s maintenance, as well as “Las Palmas”, “Hespérides” and “Sarmiento de Gamboa” Spanish Research Vessels to support the logistical operations along the 2017-2023 period. Authors also thanks to Spanish and Portuguese colleagues that collaborated in the stations and instruments’ installation, maintenance and data collection.

Financial support: Government of Spain, Ministry of Economy and Competitivity, Spanish Research Agency, by research project PERMASNOW (grant number CTM2014-52021-R), and Spanish Polar Committee by contracts PERMATHERMAL 2015, 2016, 2017, 2018, 2019, 2020, 2021, and 2022 between Geological Spanish Institute (2015-2019) and Unit of Marine Technology-National Spanish Research Council (2020-2022). Also research project FIRN (PID2022-140690OA-I00) Proyectos de Generación de Conocimiento 2022. Ministry of Science and Innovation.

\bibliographystyle{elsarticle-harv} 
\bibliography{biblio}






\end{document}